\documentclass[12pt]{article}

\usepackage{amsmath}
\usepackage{amsthm}
\usepackage{amssymb}
\usepackage{fancyvrb}
\usepackage{graphicx}
\usepackage{hyperref}
\usepackage{longtable}
\usepackage{multirow}
\usepackage{rotating}
\usepackage{color}
\usepackage[table,xcdraw]{xcolor}
\usepackage{caption}

\usepackage{fancyhdr}
\usepackage{lipsum}

\newtheorem{Th}{Theorem}
\newtheorem{Cr}{Corollary}
\newtheorem{MA}{Modelling Assumption}

\newcommand{\E}{\mathbb{E}}

\def\cpp{C\kern-.8pt{\footnotesize\raisebox{1.1pt}{+\kern-2pt+}}}

\makeatletter
\newcommand{\verbatimfont}[1]{\renewcommand{\verbatim@font}{\ttfamily#1}}
\makeatother

\definecolor{lightgray}{gray}{0.75}

\newcommand\copyrightnote[1]{}
\copyrightnote{Copyright notice}

\title{Berms without Calibration}
\author{K. E. Feldman\footnote{This paper is a personal view of the author and does not represent the view of Scotiabank. This paper is not advice. Scotiabank shall not be liable in any manner whatsoever for any consequences of loss (including but not limited to any direct, indirect or consequential loss, loss of profits and damages) arising from any reliance on or usage of this paper and accepts no legal responsibility to any party who directly or indirectly receives this material.}}
\date{Scotiabank: Innovation Hub}

\begin{document}
\maketitle

\begin{abstract}
We propose a new semi-analytical pricing model for Bermudan swaptions based on swap rate distributions and  correlations between them.
The model does not require product specific calibration.
\end{abstract}

\section*{Introduction}
A Bermudan swaption on an interest rate swap gives the holder the right to enter the swap on some specified exercise dates. This product is considered to be the simplest exotic which requires the use of a term structure model for valuations and risk management. It is customary to use an LGM style model to price Bermudan swaptions. The volatility curve of the underlying Hull-White process is calibrated to the implied volatilities of the coterminal swaptions; and the mean reversion curve is often recalibrated to the results of the Totem submissions. It is known that the mean reversion parameter is related to the forward volatility of the swaption par rates (see for example~\cite{AP2}), but calibration techniques with forward volatilities are not widely used, as it is difficult to express the forward volatility in terms of products that are liquidly traded in the interest rate market. Often midcurve swaptions (swaptions with a significant delay between the swaption expiry date and the underlying start date) are considered as products on the forward volatility of the underlying (forward starting) swap rate, though fitting them in the term structure model calibration routine is not an easy task.

In this paper we connect valuations of  Bermudan swaptions and midcurve swaptions directly. As a connecting bridge, we look at a lesser known modification of the midcurve swaption - the swaption with a relative strike where the strike is fixed sometime before the swaption exercise, but the swaption exercise is standard and within few business days of the underlying start. This swaption contract directly depends on the forward volatility between the strike fixing date and the swaption exercise date. To clarify the midcurve-Bermudan swaption relationship we study first the Canary swaption~\cite{He}, which is the simplest case of a Bermudan swaption - the swaption with only two exercise dates. Representing a Canary swaption as a basket of two option style products on  coterminal swap rates, we derive an analytic valuation based on minor assumptions on the dynamics of the swap rates between the first and the second expiry. We generalise  the valuation  to the case of arbitrary Bermudan swaption by developing a Hagan LGM~\cite{H} style model.

The main result of the present paper is similar in nature to the classical reduction of a European swaption valuation to the product of the swap annuity and a vanilla option price. We show how the numeraire change from the zero coupon bond to the annuity simplifies valuations of Bermudan swaptions by replacing  mean reversion parameters with swap rate correlations. The latter  can be easily estimated from historical data, or directly implied from midcurve swaption or CMS spread markets. 
       
The model developed in the present paper can be used to improve risk management of Bermudans, as in addition to vanilla swaptions, it allows inclusion of the simplest IR correlation products into the hedging portfolio. Another advantage of the approach is that it does not require any preliminary product specific calibration.  We also provide a toy model under the assumption of deterministic ratios of annuities and perfect correlations of the forward swap rates. In this setting we show how to value a Bermudan swaption as a zero floor on the maximum of correlated Gaussian variables using techniques presented in~\cite{SZ}. 

The paper has the following structure. In Section 1, we introduce the notations to be used throughout the paper, explain the pricing mechanism for midcurve swaptions and formulate the modelling assumption on annuity ratios.  In 
Section 2, we introduce a modification to the  swaption contract - swaptions with relative strikes, and provide its valuation formula. In Section 3, we present explicit valuation formulas for Canary swaptions under the assumption of deterministic ratios of annuities. In Section 4, we discuss an extension of the Canary swaption pricing approach to the case of a generic Bermudan swaption. We provide a general valuation mechanism and give two practical specialisations - one suitable for the moment matching valuation, and another for Hagan lattice valuation. We discuss how stochastic annuity ratios can be handled in Section 5. We summarise our findings in Conclusion.

\section{Midcurve Swaptions}
This section follows~\cite{KF} closely, and we refer the reader to it for the main details, notations and proofs. We will be working in the annuity probability measure ${\cal A}$ where the relevant underlying swap annuity is a martingale. We assume that the corresponding sample space of continuous outcomes has a time dependent filtration $\{ {\cal F}_s,s\in [t,T_e]\}$ with ${\cal F}_{s_1}\subseteq {\cal F}_{s_2}$, whenever $s_1\le s_2$,  $t$ being the time corresponding to the valuation date and $T_e$ corresponding to the end date of the underlying swap (see~\cite{AP1} for details of the use of martingale theory for interest rate modelling).
  
 A midcurve swaption is a swaption $W(t,T_{ex},S(T_s,T_e,K))$ on a forward starting swap $S(T_s,T_e,K)$ with a significant delay between the swaption expiry $T_{ex}$ and the underlying swap start date $T_s$ ($T_{ex} <T_s$, $K$ is the swaption strike which coincides with the underlying swap fixed rate). The underlying (forward starting) swap $S(T_s,T_e,K)$ can be decomposed in to the difference of the long swap $S(T_{ex},T_e,K)$ from the expiry date $T_{ex}$ to the underlying swap end date $T_e$ and the short swap $S(T_{ex},T_s,K)$ from the expiry date $T_{ex}$ to the underlying swap start date $T_s$. This decomposition allows  a midcurve swaption to be priced  in the underlying swap annuity measure as an option on a correlated weighted basket of the long swap par rate $R(T,T_{ex},T_e)$ and the short swap par rate  $R(T,T_{ex}, T_s)$ with the weights given by the ratios of the respective swap annuities - the long swap annuity $A(T,T_{ex},T_e)$ and the short swap annuity $A(T,T_{ex},T_s)$ to the underlying swap annuity $A(T,T_s,T_e)$ (see~\cite{KF} for more detailed explanation):
\begin{eqnarray}
W(t,T_{ex},S(T_s,T_e,K)) &=& A(t,T_s,T_e)\E^{\cal A}\Bigl[\Bigl(\omega(R(T_{ex},T_s,T_e) - K)\Bigr)^+\Big|{\cal F}_t\Bigr]\nonumber\\
&=& A(t,T_s,T_e)\E^{\cal A}\Bigr[\Bigl(\omega\Bigl(\frac{A(T_{ex},T_{ex},T_e)}{A(T_{ex},T_{s},T_e)}\Bigl(R(T_{ex},T_{ex},T_e) - K\Bigr)- \nonumber\\
&&\frac{A(T_{ex},T_{ex},T_s)}{A(T_{ex},T_{s},T_e)}\Bigl(R(T_{ex},T_{ex},T_s) - K\Bigr)\Bigr)^+\Big|{\cal F}_t\Bigr].
\label{MCP}
\end{eqnarray}
Here and in what follows $\omega=\pm 1$ is used to denote whether the option is a payer or a receiver. We shall use the first argument in $A(\cdot,T_s,T_e)$ and $R(\cdot,T_s,T_e)$ to specify the observation (or the fixing) time of the stochastic variable. We reserve the low case $t$ to specify the valuation date so that $A(t,T_s,T_e)$ and $R(t,T_s,T_e)$ are the expected (fair) values of the swap annuity and the swap par rate at the valuation date $t$. The upper case $T$ will be used to indicate that the variables  $A(T,T_s,T_e)$, $R(T,T_s,T_e)$ are stochastic when seen from the valuation date $t$.
As both $A(T,T_s,T_e)$, $R(T,T_s,T_e)$ are martingales in the underlying (forward starting) swap annuity measure $\cal A$ we can write:
\begin{eqnarray}
A(t,T_s,T_e) = \E^{\cal A}[A(T,T_s,T_e)],&&R(t,T_s,T_e) = \E^{\cal A}[R(T,T_s,T_e)].
\end{eqnarray}
More generally we also have
\begin{eqnarray}
A(T_{fix},T_s,T_e) = \E^{\cal A}[A(T,T_s,T_e)|{\cal F}_{T_{fix}}],&&R(T_{fix},T_s,T_e) = \E^{\cal A}[R(T,T_s,T_e)|{\cal F}_{T_{fix}}].
\end{eqnarray}

The expectation calculation in the midcurve payoff~(\ref{MCP}) can be performed by the copula integration of the payoff along the joint distribution of the long and the short swap rates in the forward starting annuity measure (see \cite{KF} for details).
If the annuity ratios $\frac{A(T_{ex},T_{ex},T_e)}{A(T_{ex},T_{s},T_e)}$ and 
$\frac{A(T_{ex},T_{ex},T_s)}{A(T_{ex},T_{s},T_e)}$ are stochastic then with the change of variables
\begin{eqnarray}
\tilde x & = & (x-K)\frac{A(t,T_{s},T_e)}{A(t,T_{ex},T_e)} \E^{\cal A}\left[\frac{A(T_{ex},T_{ex},T_e)}{A(T_{ex},T_{s},T_e)}\Big|x=R(T_{ex},T_{ex},T_e), y=R(T_{ex},T_{ex},T_s),{\cal F}_{T_{ex}}\right]  \nonumber\\
&=& (x-K)f(x,y),\nonumber\\
\tilde y & = & (y-K)\frac{A(t,T_{s},T_e)}{A(t,T_{ex},T_s)}\E^{\cal A}\left[\frac{A(T_{ex},T_{ex},T_s)}{A(T_{ex},T_{s},T_e)} \Big|x=R(T_{ex},T_{ex},T_e), y=R(T_{ex},T_{ex},T_s),{\cal F}_{T_{ex}}\right] \nonumber\\
&=&(y-K)g(x,y),\nonumber
\end{eqnarray}
the copula integration is reduced to 
\begin{eqnarray}
W(t,T_{ex},S(T_s,T_e,K)) &=& A(t,T_s,T_e)\int^{\infty}_{-\infty}\int^{\infty}_{-\infty}(a\tilde x- b\tilde y)^+
\Psi(\Psi^u_e(x),\Psi^u_s(y))pdf^u_e(x)pdf^u_s(y)\times\nonumber\\
&&det\left(
\begin{array}{cc}
\frac{\partial \tilde x}{\partial x}& \frac{\partial \tilde y}{\partial x}\\
\frac{\partial \tilde x}{\partial y}& \frac{\partial \tilde y}{\partial y}
\end{array}
\right) ^{-1}d\tilde x d \tilde y,\nonumber\\
&&a=\frac{A(t, T_{ex},T_e)}{A(t,T_{s},T_e)}, \quad b=\frac{A(t,T_{ex},T_s)}{A(t,T_{s},T_e)},
\label{GC}
\end{eqnarray}
where $\Psi$ is the copula kernel and $\Psi^u_e,\Psi^u_s$ are the cumulative density functions of the swap rates distributions $pdf^u_e, pdf^u_s$ in the underlying (forward starting) swap annuity measures. The integration weight can be simplified further:
\begin{eqnarray}
det\left(
\begin{array}{cc}
\frac{\partial \tilde x}{\partial x}& \frac{\partial \tilde y}{\partial x}\\
\frac{\partial \tilde x}{\partial y}& \frac{\partial \tilde y}{\partial y}
\end{array}
\right)^{-1}&=&det\left(
\begin{array}{cc}
f+(x-K)f_x& (y-K)g_x\\
(x-K)f_y& g+(y-K)g_y
\end{array}\right)^{-1}\nonumber\\
&=&f(\cdot)^{-1}g(\cdot)^{-1}\left(1 +(x-K)\frac{f_x}{f} +(y-K)\frac{g_y}{g}+(x-K)(y-K)\frac{f_xg_y-f_yg_x}{fg}\right)^{-1}\nonumber\\
&=&f(\cdot)^{-1}g(\cdot)^{-1}\left(1 +(x-K)\frac{f_x}{f} + (y-K)\frac{g_y}{g}\right)^{-1},
\end{eqnarray}
where for the last transformation we used that $af(x,y) =  bg(x,y) + 1$, which is a consequence of the following identity:
\begin{eqnarray}
\frac{A(T_{ex},T_{ex},T_e)}{A(T_{ex},T_{s},T_e)} - \frac{A(T_{ex},T_{ex},T_s)}{A(T_{ex},T_{s},T_e)} &\equiv& 1.
\end{eqnarray}
If we assume that annuity ratios $\frac{A(T,T_{ex},T_e)}{A(T,T_{s},T_e)}$ and $\frac{A(T,T_{ex},T_s)}{A(T,T_{s},T_e)}$ are deterministic, then the weight factor $fg\left(1 + (x-K)\frac{f_x}{f} + (y-K)\frac{g_y}{g}\right)$ is 1. In the general case, it is convenient to linearise the weight in such a way that the following assumption holds:
\begin{MA} We  assume that there exist such deformations of the probability densities $pdf^u_e(x)$ and $pdf^u_s(y)$ that
\begin{eqnarray}
\frac{\Psi(\Psi^u_e(x),\Psi^u_s(y)) pdf^u_e(x)pdf^u_s(y)}{f(x,y)g(x,y)\left(1 + (x-K)\frac{f_x(x,y)}{f(x,y)} + (y-K)\frac{g_y(x,y)}{g(x,y)}\right)}\approx\Psi(\widetilde{\Psi_e}(\tilde x),\widetilde{\Psi_s}(\tilde y))\widetilde {pdf}_e(\tilde x)\widetilde {pdf}_s(\tilde y)\nonumber\\
\label{LNZ}
\end{eqnarray}
for the deformed probability densities $\widetilde {pdf}_e(\tilde x)$ and $\widetilde {pdf}_s(\tilde y)$, where by $\widetilde {\Psi_e}(\tilde x)$ and $\widetilde {\Psi_s}(\tilde y)$ we denote the cumulative density functions of the deformed distributions. 
\end{MA}
\noindent
{\bf Example:} If the initial swap rates distributions are Gaussian and the annuity ratios are log linear  in the long and the short swap rates as in Section 2 of~\cite{KF}
then in the Gaussian copula one can use (with an appropriate choice of parameters $\mu_e$, $\mu_s$, $\delta_e$ and $\delta_s$)
\begin{eqnarray}
\widetilde {pdf}_e(\tilde x)&=& e^{\delta_e(\tilde x - \mu_e)}pdf_e(\tilde x)=\frac{1}{\sqrt{2\pi}\sigma_e}
e^{\delta_e(\tilde x - \mu_e)}e^{-\frac{(\tilde x-\tilde x_e)^2}{2\sigma^2_e}}=\frac{1}{\sqrt{2\pi }\sigma_e}e^{-\frac{(\tilde x-\tilde x_e -\sigma^2_e\delta_e)^2}{2\sigma^2_e}},
\nonumber\\
\widetilde {pdf}_s(\tilde y)&=&e^{\delta_s(\tilde y - \mu_s)}pdf_s(\tilde y)=\frac{1}{\sqrt{2\pi }\sigma_s}
e^{\delta_s(\tilde y - \mu_s)}e^{-\frac{(\tilde y-\tilde y_s)^2}{2\sigma^2_s}}=\frac{1}{\sqrt{2\pi }\sigma_s}e^{-\frac{(\tilde y-\tilde y_s -\sigma^2_s\delta_s )^2}{2\sigma^2_s }},
\end{eqnarray}
i.e. the deformations amount to convexity adjustments on the respective forwards ($\sigma_e$, $\sigma_s$  are the implied volatilities of the long and the short swap rates respectively inclusive the square root time factor).

\

\noindent
In the following sections we will rely on the following consequences of~\cite{KF} and linearisation~(\ref{LNZ}):

\begin{Th}
Under the assumption of deterministic ratios of annuities, the Radom-Nikodym derivative of the measure changes when moving between the long, the short and the forward starting swap annuity measures are all equal to 1. Thus, we can use distributions of the corresponding swap rates as observed in their natural measures and substitute them to the joint distribution of those rates in each of the measure.
\end{Th}
\noindent
{\bf Proof:} Using Lemma 1 from \cite{KF} we can relate the probability densities $pdf_e(x)$, $pdf_s(y)$  of the long and the short swap rates in their natural annuities measures to the probability densities $pdf^u_e(x)$, $pdf^u_s(y)$ of the same swap rates in the underlying (forward starting) swap annuity measure as:
\begin{eqnarray}
pdf^u_e(x)&=& pdf_e(x)\frac{A(t,T_{ex},T_e)}{A(t,T_s,T_e)}\E^{{\cal A}_e}\left[\frac{A(T_{ex},T_s,T_e)}{A(T_{ex},T_{ex},T_e)}\Big|x=R(T_{ex},T_{ex},T_e),{\cal F}_{T_{ex}}\right],\nonumber\\
pdf^u_s(y)&=& pdf_s(y)\frac{A(t,T_{ex},T_s)}{A(t,T_s,T_e)}\E^{{\cal A}_s}\left[\frac{A(T_{ex},T_s,T_e)}{A(T_{ex},T_{ex},T_s)}\Big|y=R(T_{ex},T_{ex},T_s),{\cal F}_{T_{ex}}\right],
\end{eqnarray}
where ${\cal A}_e$ and ${\cal A}_s$ are the long and the short swap annuity measures respectively.
If the annuity ratios are deterministic then the terms under expectations cancel the normalising coefficients in front of them. This gives the result of the theorem.

\begin{Th}
\label{TLNZ}
Under the linearisation assumption~(\ref{LNZ}) we can make convexity adjustments to the relevant swap rates so that payoffs expressed as functions of 
\begin{eqnarray}
\frac{A(T_{ex},T_{ex},T_e)}{A(T_{ex},T_{s},T_e)}(R(T_{ex},T_{ex},T_e)-K)&{\rm and} &\frac{A(T_{ex},T_{ex},T_s)}{A(T_{ex},T_{s},T_e)}(R(T_{ex},T_{ex},T_s)-K)
\end{eqnarray}
 can be valued using  the deterministic annuity ratio assumption. 
\end{Th}
\noindent
{\bf Proof:} If we make a measure change from the underlying (forward starting) swap annuity measure to the measure implied by the joint distribution of the deformations $\widetilde {pdf}_e(\tilde x)$ and $\widetilde {pdf}_s(\tilde y)$ from the linearisation assumption, then we can price any payoff which is a function of $\frac{A(T,T_{ex},T_e)}{A(T,T_{s},T_e)}(R(T,T_{ex},T_e)-K)$ and $\frac{A(T,T_{ex},T_s)}{A(T,T_{s},T_e)}(R(T,T_{ex},T_s)-K)$ by integrating a function of $a\tilde x$ and $b\tilde y$ along the joint distribuiton $\widetilde {pdf}_e(\tilde x)$ and $\widetilde {pdf}_s(\tilde y)$ ($a=A(t,T_{ex},T_e)/A(t,T_{s},T_e)$, $b=A(t,T_{ex},T_s)/A(t,T_{s},T_e)$). Thus, the result of the valuation will be the same as if we assume deterministic annuity ratios with additional convexity adjustments to the long and the short swap rate distributions.

\

 Practitioners often manage midcurve swaptions in terms of the midcurve implied volatility-by-strike which is approximated as
\begin{eqnarray}
\sigma^2_{s,e}(K)&=&\frac{A(t,T_{ex},T_e)^2}{A(t,T_{s},T_e)^2}\sigma^2_e(K_e) - 2\frac{A(t,T_{ex},T_e)}{A(t,T_{s},T_e)}\frac{A(t,T_{ex},T_s)}{A(t,T_{s},T_e)}\rho_{s,e}\sigma_e(K_e)\sigma_s(K_s) + \frac{A(t,T_{ex},T_s)^2}{A(t,T_{s},T_e)^2}\sigma^2_s(K_s),\nonumber\\
\end{eqnarray}
where $K_e = R(t,T_{ex},T_e) + (K- R(t,T_{s},T_e))$, $K_s = R(t,T_{ex},T_s) + (K- R(t,T_{s},T_e))$ and $\rho_{s,e}$ is the implied correlation of the long and short swap rates as fixed at $T_{ex}$. We shall use this approximation (omitting the strike $K$ from the notations) to present results in the commonly used terminology of implied volatilities, while the procedure can be made rigorous by switching to local volatilities and using Andreasen-Huge one-time-step method~\cite{AH}, or using Black/SABR baskets approach from~\cite{AA,FKM,HLSW}.

\newpage

\section{Swaptions with Relative Strikes}

When a vanilla swaption is traded, its strike is typically  quoted as a spread relative to the prevailing ATM level - the par rate of the underlying swap. Once the swaption contract is agreed and  entered by the counterparties, the product is booked with an absolute strike equal to the  sum of ATM level and the spread. In this section we consider a modification to the swaption contract where the ATM level is also a part of the contract and its fixing time $T_{fix}$  can be any time between the contract issue date and the swaption exercise date $T_{ex}$.  

First, consider the case when the ATM level -  the par rate of the underlying swap  - is fixed at expiry $T_{ex}$, i.e. the absolute strike of the swaption becomes only known/fixed at the swaption expiry. This  payoff is deterministic
\begin{eqnarray}
W(t,T_{ex},S(T_s,T_e,\E^{\cal A}[R(T_{s},T_s,T_e)|{\cal F}_{T_{ex}}]+K)) &=& A(t,T_s,T_e)\E^{\cal A}\Bigl[\Bigl(\omega(R(T_{ex},T_s,T_e) - \nonumber\\
&&\E^{\cal A}[R(T_{s},T_s,T_e)|{\cal F}_{T_{ex}}]-K)\Bigl)^+\Bigl|{\cal F}_t\Bigl]\nonumber\\
&=&A(t,T_s,T_e)(\omega K)^+.
\end{eqnarray}
Assume instead that  the swaption expiry $T_{ex}$ matches the underlying swap start date $T_s$ ($T_{ex}=T_s$) but the ATM level is fixed at $T_{fix}<T_s$, then:
\begin{eqnarray}
W(t,T_s,S(T_s,T_e,\E^{\cal A}[R(T_s, T_s,T_e)|{\cal F}_{T_{fix}}]+K)) &=& A(t,T_s,T_e)\E^{\cal A}\Bigr[\Bigr(\omega(R(T_s,T_s,T_e) -\nonumber\\
&& \E^{\cal A}[R(T_{s},T_s,T_e)|{\cal F}_{T_{fix}}]-K)\Bigl)^+\Bigl|{\cal F}_t\Bigl].
\end{eqnarray}
The expectation above is with respect to the joint distribution of the swap rate $R(T_s,T_s,T_e)$ as fixed at $T_s$ and the swap rate $R(T_{fix},T_s,T_e)$ as fixed at $T_{fix}$ and can be performed by copula integration.
The probability density for the swap rate $R(T_s,T_s,T_e)$ as fixed at $T_s$ is the standard distribution observed in the vanilla swaption market.  The distribution of the swap rate $R(T_{fix},T_s,T_e)$ as fixed at $T_{fix}$  is the midcurve swaption rate distribution modelled  as a joint distribution of $R(T_{fix},T_{fix},T_e)$ as fixed at $T_{fix}$ and $R(T_{fix},T_{fix},T_s)$ as fixed at $T_{fix}$, both  are the standard swap rate distributions observed in the vanilla swaption market.
To use the copula integration, one will need to estimate the correlation 
\begin{eqnarray}
\rho = \rho(T_{fix},T_s, R(T,T_s,T_e)) &=& corr(R(T_s,T_s,T_e), R(T_{fix},T_s,T_e)).
\end{eqnarray}
\begin{Th}
\label{MRSCorr}
The correlation $\rho$ between the swap rates $R(T,T_s,T_e)$ as fixed at $T_{fix}$ and as fixed at $T_s$ ($T_{fix}<T_s$) is: 
\begin{eqnarray}
\rho &=&\frac{vol(R(T_{fix},T_s,T_e))\sqrt{T_{fix}}}{vol(R(T_s,T_s,T_e))\sqrt{T_s}}. 
\end{eqnarray}
\end{Th}
\noindent
{\bf Proof: } Applying the tower rule we obtain
\begin{eqnarray}
\rho &=& \frac{\E^{\cal A}[R(T_s,T_s,T_e) \E^{\cal A}[R(T_s,T_s,T_e)|{\cal F}_{T_{fix}}]]-\E^{\cal A}[R(T_s,T_s,T_e)]^2}{vol(R(T_s,T_s,T_e))vol(R(T_{fix},T_s,T_e))\sqrt{T_{fix}T_s}}\nonumber\\
 &=& \frac{\E^{\cal A}[\E^{\cal A}[R(T_s,T_s,T_e)|{\cal F}_{T_{fix}}]^2]-\E^{\cal A}[R(T_s,T_s,T_e)]^2}{vol(R(T_s,T_s,T_e))vol(R(T_{fix}T_s,T_e))\sqrt{T_{fix}T_s}}\nonumber\\
 &=& \frac{vol(R(T_{fix},T_s,T_e))^2T_{fix}}{vol(R(T_{s},T_s,T_e))vol(R(T_{fix},T_s,T_e))\sqrt{T_{fix}T_s}}\nonumber\\
&=&\frac{vol(R(T_{fix},T_s,T_e))\sqrt{T_{fix}}}{vol(R(T_{s},T_s,T_e))\sqrt{T_s}}.
\end{eqnarray}

\subsubsection*{Swaptions with Relative Strikes: Gaussian Case}
Let us assume that all the standard swap rate distributions above are  Gaussian and denote
\begin{eqnarray}
\sigma_z&=&vol(R(T_{fix},T_s,T_e))\sqrt{T_{fix}},\nonumber\\
\sigma_x&=&vol(R(T_{s},T_s,T_e))\sqrt{T_s},
\end{eqnarray}
so that $\rho=\sigma_z/\sigma_x$.
\begin{Th}  Under Gaussian assumptions the fair value of a swaption with a relative strike $K$ is:
\begin{eqnarray} 
W(t,T_s,S(T_s,T_e,\E^{\cal A}[R(T_s,T_s,T_e)|{\cal F}_{T_{fix}}]+K)) &=&A(t,T_s,T_e)\Omega(0,K,\sqrt{\sigma^2_x-\sigma^2_z},\omega),
\end{eqnarray}
where $\Omega(F,K,\sigma,\omega)$ is the undiscounted Bachelier option price:
\begin{eqnarray}
\Omega(F,K,\sigma,\omega)&=&\omega(F - K)\Phi\left(\frac{\omega(F - K)}{\sigma}\right) + \sigma\phi\left(\frac{F - K}{\sigma}\right).
\end{eqnarray}
\label{MRS}
\end{Th}
\noindent
{\bf Proof:}
The payoff $W(t,T_s,S(T_s,T_e,\E^{\cal A}[R(T_s,T_s,T_e)|{\cal F}_{T_{fix}}]+K))$ can be computed by integration:
\begin{eqnarray}
\frac{W(t,T_s,S(T_s,T_e,\E^{\cal A}[R(T_s,T_s,T_e)|{\cal F}_{T_{fix}}]+K))}{A(t,T_s,T_e)}&=&\int^{\infty}_{-\infty}\int^{\infty}_{-\infty}\frac{ (\omega(x-z-K))^+}{2\pi\sigma_x\sigma_z\sqrt{1-\rho^2}}e^{-\frac{\sigma^2_zx^2-2\rho \sigma_x\sigma_zx z +\sigma^2_xz^2}{2(1-\rho^2)\sigma^2_x\sigma^2_z}}dxdz\nonumber\\
&=&\int^{\infty}_{-\infty}\int^{\infty}_{-\infty}\frac{ (\omega(x-z-K))^+}{2\pi\sigma_x\sigma_z\sqrt{1-\rho^2}}e^{-\frac{\sigma^2_z(x-z)^2+(\sigma^2_x -\sigma^2_z)z^2}{2(1-\rho^2)\sigma^2_x\sigma^2_z}}dxdz\nonumber\\
&=&\int^{\infty}_{-\infty}\frac{e^{-\frac{z^2}{2\sigma^2_z}}}{\sqrt{2\pi}\sigma_z}dz\int^{\infty}_{-\infty}\frac{ (\omega(t-K))^+e^{-\frac{t^2}{2(\sigma^2_x-\sigma^2_z)}}}{\sqrt{2\pi(\sigma^2_x-\sigma^2_z)}}dt\nonumber\\
&=&\int^{\infty}_{-\infty}\frac{ (\omega(t-K))^+e^{-\frac{t^2}{2(\sigma^2_x-\sigma^2_z)}}}{\sqrt{2\pi(\sigma^2_x-\sigma^2_z)}}dt\nonumber\\
&=&\Omega(0,K,\sqrt{\sigma^2_x-\sigma^2_z},\omega).
\end{eqnarray}

 We also note that in the case $\sigma_x=\sigma_z$ the payoff of the relative strike swaption yet again becomes deterministic and is 
$A(t,T_s,T_e)(\omega K)^+$. The case $\sigma_x=\sigma_z$  corresponds to the choice of the correlation between the long and the short swap rates, both fixed at $T_{fix}$, as:
\begin{eqnarray}
\rho_{s,e}=\frac{A(t,T_{fix},T_e)^2\sigma^2_e  +  A(t,T_{fix},T_s)^2\sigma^2_s - A(t,T_s,T_e)^2 \sigma^2_x}{2A(t,T_{fix},T_e)  A(t,T_{fix},T_s)\sigma_e\sigma_s}.
\end{eqnarray}

\subsubsection*{Swaptions with Relative Strikes: General Case}

As there is no developed market for the relative strike swaptions, the adequate management of the product requires a dedicated parameter. To preserve the relation with midcurves, the parameter can be managed as a multiplicative spread on top of the relative strike swaption volatility $\sqrt{\sigma^2_x-\sigma^2_z}$. To  not burden the explanation, we shall not use the parameter in the following sections, as all the derivations can be easily adjusted.

\section{Canary Swaptions}
A Canary swaption~\cite{He} is the simplest Bermudan swaption with only two exercise dates: $T_1<T_2$. There are two coterminal swaps the swaption can be exercised into: $S(T_1,T_e,K)$ and $S(T_2,T_e,K)$. 
The payoff can be calculated in the annuity measure of the second coterminal swap as
\begin{eqnarray}
&&\frac{C(t,T_1,T_2,T_e,K)}{A(t,T_2,T_e)} =\nonumber\\
&&\E^{{\cal A}_2}\Bigl[max\Bigl\{\frac{A(T_1,T_1,T_e)}{A(T_1,T_2,T_e)}\Bigl(\omega(R(T_1,T_1,T_e)-K)\Bigr)^+,\E^{{\cal A}_2}\Bigl[\Bigl(\omega(R(T_2,T_2,T_e)-K)\Bigr)^+|{\cal F}_{T_1}\Bigr]\Bigr\}\Big|{\cal F}_t\Bigr].\nonumber\\
\label{CSP}
\end{eqnarray}

To evaluate the overall expectation  the stochastic variable $(\omega(R(T_1,T_1,T_e)-K))^+$ needs to be moved from its natural ${\cal A}_1$ annuity measure, which corresponds to the annuity $A(T,T_1,T_e)$ of the first coterminal swap $S(T_1,T_e,K)$,  to ${\cal A}_2$ annuity measure corresponding to the annuity $A(T,T_2,T_e)$ of the second cotermianl swap. 
The change of measure is the forward starting swaption $W(T_1,S(T_2,T_e,K))$ measure change constructed in Section 1. In this section we will assume that the annuity ratio $\frac{A(T_1,T_1,T_e)}{A(T_1,T_2,T_e)}$ is deterministic. In the general case the Canary swaption payoff~(\ref{CSP}) can be treated within Theorem~\ref{TLNZ}   as the payoff formula is a function of $\frac{A(T_1,T_{1},T_e)}{A(T_1,T_{2},T_e)}(R(T_1,T_{1},T_e)-K)$ and $\frac{A(T_1,T_{1},T_2)}{A(T_1,T_{2},T_e)}(R(T_1,T_{1},T_2)-K)$.

For the second expectation in~(\ref{CSP}), i.e. the one inside the curly brackets and corresponding to $R(T_2,T_2,T_e)$, we can write:
\begin{eqnarray}
\E^{{\cal A}_2}\Bigl[\Bigl(\omega(R(T_2,T_2,T_e)-K)\Bigr)^+|{\cal F}_{T_1}\Bigr] &=&\E^{{\cal A}_2}\Bigl[\Bigl(\omega(R(T_2,T_2,T_e)-\E^{{\cal A}_2}[R(T_2,T_2,T_e)|{\cal F}_{T_1}] +\nonumber\\
&& \E^{{\cal A}_2}[R(T_2,T_2,T_e)|T_1] - K)\Bigr)^+\Big|{\cal F}_{T_1}\Bigr]\nonumber\\
&=&\E^{{\cal A}_2}\Bigl[\left(\omega(N\left(0,\sqrt{\sigma^2_{X_2}-\sigma^2_{Z}}\right) + \E^{{\cal A}_2}[R(T_2,T_2,T_e)|{\cal F}_{T_1}] - K)\right)^+\Big|{\cal F}_{T_1}\Bigr]\nonumber\\
&=&\E^{{\cal A}_2}\bigl[\Omega\left(0,K-z,\sqrt{\sigma^2_{X_2}-\sigma^2_{Z}},\omega\right)\Big|z=\E^{{\cal A}_2}[R(T_2,T_2,T_e)|{\cal F}_{T_1}],{\cal F}_{T_1}\Bigr],\nonumber\\
\label{SecExp}
\end{eqnarray}
where $\sigma_{X_2}$ is the implied volatility at strike (inclusive the square root time factor) of the distribution $pdf_{X_2}$ for $X_2=R(T_2,T_2,T_e)$ (i.e. as observed at $T_2$) and $\sigma_Z$ is the implied volatility at strike (inclusive the square root time factor) of the distribution $pdf_Z$ for $Z=R(T_1,T_2,T_e)$ (i.e. as observed at $T_1$); $N(\mu,\sigma)$ is used to denote the relevant normal random variable with the mean $\mu$ and the variance $\sigma^2$, $\Omega(\cdot)$ is the Bachelier option price as described in Section 2. Note that in~(\ref{SecExp}) we are using references to the normal distribution and Bachelier option formula for the convenience of the explanation. In general~(\ref{SecExp}) 
is the option part of the payoff of a swaption with the relative strike as described in Section 2. One can use a multiplicative spread over  $\sqrt{\sigma^2_{X_2}-\sigma^2_{Z}}$ or a copula integration with the correlation from Theorem~\ref{MRSCorr} to express the dependence on the volatility smile.

\subsubsection*{Canary Swaptions: Gaussian Case}
Let us assume that all the standard swap rate distributions above are  Gaussian. 

\begin{Th}\label{CSPTh}  Assume that the joint distribution of $pdf_{X_1}$ and $pdf_Z$ is 
Gaussian with the correlation $\rho_{X_1,Z}$ ($\rho_{X_1,Z}$ is the correlation between the swap rate $R(T_1,T_1,T_e)$ and the midcurve underlying swap rate $R(T_1,T_2,T_e)$).
The price of a Canary swaption is:
\begin{eqnarray}
\frac{C(t,T_1,T_2,T_e,K)}{A(t,T_2,T_e)}&=&\int^{\infty}_{-\infty}\int^{\infty}_{-\infty}max\Bigl \{
\frac{A(t,T_1,T_e)}{A(t,T_2,T_e)}
\left(\omega(x-K)\right)^+,
\Omega\left(0,K-z,
\sqrt{\sigma^2_{X_2}-\sigma^2_Z},\omega\right)\Bigr\}\nonumber\\
&\times&e^{-\frac{(x-\mu_1)^2-2\rho_{X_1,Z}\frac{\sigma_{1e}}{\sigma_Z}(x-\mu_1)(z-\mu_2)+\frac{\sigma^2_{1e}}{\sigma^2_Z}(z-\mu_2)^2}{2(1-\rho^2_{X_1,Z})\sigma^2_{1e}}}\frac{dxdz}{2\pi \sigma_{1e}\sigma_Z\sqrt{1-\rho^2_{X_1,Z}}},
\label{CSWPNF}
\end{eqnarray}
where  $\mu_1=R(t,T_1,T_e)$, $\sigma_{1e}$ is the volatility of $R(T_1,T_1,T_e)$,  $\mu_{2}=R(t,T_2,T_e)$ and $\sigma_Z$ is the midcurve volatility of $R(T_1,T_2,T_e)$.
\end{Th}
\noindent
{\bf Proof: } With the substitution~(\ref{SecExp}) the expectation~(\ref{CSP}) becomes an integral over a joint distribution of $pdf_{X_1}$ and $pdf_Z$. Under the Gaussian assumption the joint distribution is the standard 2d-normal distribution with means, vols and the correlation as specified in the theorem.

\begin{Th}
Let in addition $\sigma_{12}$ be the volatility of the short swap rate $R(T_1,T_1,T_2)$, then 
\begin{eqnarray}
\rho_{X_1,Z}&=&\frac{A(t,T_1,T_e)\sigma_{1e} - A(t,T_1,T_2) \rho_{2,e}\sigma_{12}}{\sqrt{A(t,T_1,T_e)^2\sigma^2_{1e} -2A(t,T_1,T_e)A(t,T_1,T_2)\rho_{2,e} \sigma_{1e}\sigma_{12} +  A(t,T_1,T_2)^2\sigma^2_{12}}},
\end{eqnarray}
\end{Th}
\noindent
{\bf Proof: } The probability density $pdf_Z$ is the midcurve swap rate distribution and is described via the joint density of $pdf_{X_1}$ for $X_1=R(T_1,T_1,T_e)$  and $pdf_Y$ for $Y=R(T_1,T_1,T_2)$ both as observed at $T_1$ (see Section 1):
\begin{eqnarray}
R(T_1,T_2,T_e) &=& \frac{A(T_1,T_1,T_e)}{A(T_1,T_2,T_e)} R(T_1,T_1,T_e) - \frac{A(T_1,T_1,T_2)}{A(T_1,T_2,T_e)} R(T_1,T_1,T_2),
\end{eqnarray}
so that for the deterministic annuity ratios, the volatility of $Z=R(T_1,T_2,T_e)$ is
\begin{eqnarray}
\sigma^2_Z =\sigma^2_{2e} = \frac{A(t,T_1,T_e)^2}{A(t,T_2,T_e)^2}\sigma^2_{1e} -2\frac{A(t,T_1,T_e)A(t,T_1,T_2)}{A(t,T_2,T_e)^2}\rho_{2,e} \sigma_{1e}\sigma_{12} +  \frac{A(t,T_1,T_2)^2}{A(t,T_2,T_e)^2}\sigma^2_{12}.
\end{eqnarray}
The covariance between $Z$ and $X_1$ is
\begin{eqnarray}
Cov(X_1,Z)&=&\frac{A(t,T_1,T_e)}{A(t,T_2,T_e)}\sigma^2_{1e} - \frac{A(t,T_1,T_2)}{A(t,T_2,T_e)} \rho_{2,e}\sigma_{12}\sigma_{1e},
\end{eqnarray}
and the correlation is
\begin{eqnarray}
\rho_{X_1,Z}&=&\frac{A(t,T_1,T_e)\sigma_{1e} - A(t,T_1,T_2) \rho_{2,e}\sigma_{12}}{\sqrt{A(t,T_1,T_e)^2\sigma^2_{1e} -2A(t,T_1,T_e)A(t,T_1,T_2)\rho_{2,e} \sigma_{1e}\sigma_{12} +  A(t,T_1,T_2)^2\sigma^2_{12}}},
\end{eqnarray}
where $\rho_{2,e}$ is the midcurve correlation (i.e. between $R(T_1,T_1,T_e)$ and $R(T_1,T_1,T_2)$). 

\begin{Cr}
\label{MntC}
In the case of the perfect correlation between the forward rates $Z$ and $X_2$ we can simplify
\begin{eqnarray}
\rho_{X_1,Z}&=&\frac{A(t,T_1,T_e)^2\sigma^2_{1e} +A(t,T_2,T_e)^2\sigma^2_{2e} - A(t,T_1,T_2)^2\sigma^2_{12}}{2A(t,T_1,T_e)A(t,T_2,T_e)\sigma_{2e}\sigma_{1e}},
\end{eqnarray}
and the option (\ref{SecExp}) becomes deterministic, i.e. the intrinsic value. In this case the Canary swaption payoff (\ref{CSP}) is approximated by the maximum of the following two correlated normal variables:
\begin{eqnarray}
N\left(\frac{A(t,T_1,T_e)}{A(t,T_2,T_e)}(\omega(R(t,T_1,T_e)-K)),\frac{A(t,T_1,T_e)}{A(t,T_2,T_e)}\sigma_{1e}\right), &\quad& N\left(\omega(R(t,T_2,T_e)-K),\sigma_{12}\right).\nonumber\\
\label{MntCF}
\end{eqnarray} 
\end{Cr}
\noindent
In the next sections we will refer to the condition of the Corollary~\ref{MntC} as the condition of perfect correlations of the forward rates. This is because $Z$ and $X_2$ are distributions of the same swap rate but observed at different times in the future. 

\subsubsection*{Canary Swaptions: General Case}
Theorem~\ref{CSPTh} can be generalised to the full smile case by moving to the copula integration and actual distributions for $pdf_{X_1}$, $pdf_Z$ and $pdf_{X_2}$ (or the probability density of $\sigma_{X_2}$).

\

We performed a comparison of valuations based on Theorem~\ref{CSPTh} and Corollary~\ref{MntC} in real market conditions observed close of business day on 30  June 2025 against the classical Hagan LGM. The volatility surface used for all three models is SABR surface calibrated to the observed European swaption market. The valuation based on Theorem~\ref{CSPTh} is performed using Simpson's rule for the double integral;  the maximum in Corollary~\ref{MntC} is evaluated  by the moment matching techniques as outlined in~\cite{SZ}. The correlation between the long and the short swap rates is  89\%.  To account for the impact of the volatility smile we chose volatilities at relevant strikes when evaluating individual Bermudan prices via~(\ref{CSWPNF}) or~(\ref{MntCF}).   Figure 1 shows the difference of each of the three valuations with the market consensus for a range of available strikes between ATM-200bps and ATM+600bps for a payer callable swap, which starts in 2 years and can be exercised annually into either 2 or 1 year swap. This is the product equivalent to a Canary swaption 2Y1Y2Y described in this section.  The doted lines are plus/minus one standard deviations of the market quotes.

\begin{figure}[h!]
\centering
\includegraphics{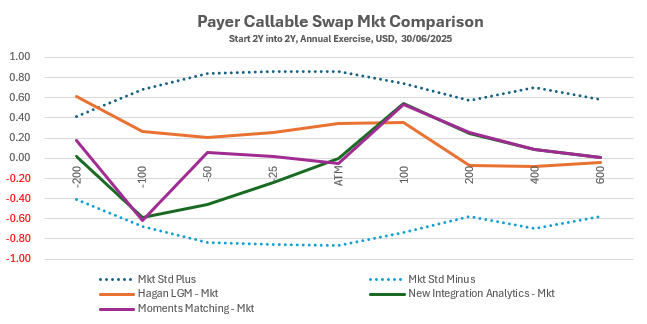}
\small{\footnotesize\caption{Model price vs market consensus under different model assumptions: the callable swap equivalent to the Canary swaption on a 2Y swap starting in 2Y with the annual exercise frequency.}}
\end{figure}

\newpage

\section{Berms}

Analytic pricing of Canary swaptions from the previous section can be extended to pricing of Bermudan swaptions by induction. Consider a Bermudan swaption $B(t,\{T_j\},T_e,K)$  with exercise times $\{T_j\}=T_1< T_2<\cdots <T_n$, $T_n<T_e$, and the strike $K$. At each exercise time $T_j$ the option holder may enter into a swap $S(T_j,T_e, K)$.  We can bring all the swap rates distribution into the same annuity measure  ${\cal A}_n$ of the last swap and evaluate the payoff as
\begin{eqnarray}
\frac{B(t,\{T_j\},T_e,K)}{A(t,T_n,T_e)} =\E^{{\cal A}_n}\Bigl [max\Bigl (\frac{A(T_1,T_1,T_e)}{A(T_1,T_n,T_e)}(\omega(x_1-K))^+,\E^{{\cal A}_n}\Bigl [max\Bigl (\frac{A(T_2,T_2,T_e)}{A(T_2,T_n,T_e)}(\omega(x_2-K))^+,\cdots\nonumber\\
\cdots,\E^{{\cal A}_n}\Bigl [max\Bigl(\frac{A(T_{n-1},T_{n-1},T_e)}{A(T_{n-1},T_n,T_e)}(\omega(x_{n-1}-K))^+,\E^{{\cal A}_n}\Bigl [(\omega(x_{n}-K))^+ |x_n=R(T_n,T_n,T_e),{\cal F}_{T_{n-1}}\Bigr]\Bigr )\Bigr | \nonumber\\
\Bigr |x_{n-1}=R(T_{n-1},T_{n-1},T_e),{\cal F}_{T_{n-2}}\Bigr]\Bigr )\cdots \Bigr |x_2 = R(T_2,T_2,T_e),{\cal F}_{T_1}\Bigr ]\Bigr )\Bigr |x_1=R(T_1,T_1,T_e),{\cal F}_t\Bigr ].\nonumber\\
\label{GENERICE}
\end{eqnarray}
The above expectation is a sequence of  integrals of the payoff functions over joint probability densities of  the swap rate distributions as observed at the respective times $T_i$, $i=1,...,n$. To evaluate the integral at each next step $i$ we need to get expectations of the swap rates $R(T_{i+1},T_{i+1},T_e),..., R(T_n,T_n,T_e)$ with respect to the observation time $T_i$ and perform
integrations over all stochastic variables of the form 
\begin{eqnarray}
\E^{{\cal A}_n}[R(T_{i+k},T_{i+k},T_e)|{\cal F}_{T_{i+1}}] - \E^{{\cal A}_n}[R(T_{i+k},T_{i+k},T_e)|{\cal F}_{T_i}].
\end{eqnarray}
The result is  a stochastic variable over a joint distribution of $\E^{{{\cal A}_n}}[R(T_{i+k},T_{i+k},T_e)|{\cal F}_{T_i}]$, $k=0,...n-i$. The distribution of the  random variable $\E^{{\cal A}_n}[R(T_{i+k},T_{i+k},T_e)|{\cal F}_{T_i}]$  is  the underlying (forward starting) swap rate distribution for a midcurve swaption on $R(T_i,T_{i+k},T_e)$ but observed in (unnatural) annuity measure ${{\cal A}_n}$. The measure change is trivial under the deterministic annuity ratio assumption while in the general case it can be handled with an appropriate convexity adjustment using Theorem~\ref{TLNZ} . To find the correlation matrix of this joint distribution 
we can use an alternative derivation for the Canary swaption payoff expectation as an integral over a joint probability density of various swap rates  as follows. Consider for any $i,j$ satisfying $1\le i< j\le n$, the  swap rate distributions

\begin{itemize}
\item $pdf_{X_i}$ for $R(T_i,T_i,T_e)$ as observed at $T_i$,
\item $pdf_{X_j}$ for $R(T_j,T_j,T_e)$ as observed at $T_j$, 
\item $pdf_Z$ for $R(T_i,T_j,T_e)$ as observed at $T_i$.
\end{itemize}
The later can be described via a joint density of $pdf_{X_i}$ for $R(T_i,T_i,T_e)$  and $pdf_Y$ for $R(T_i,T_i,T_j)$ both as observed at $T_i$ via
\begin{eqnarray}
R(T_i,T_j,T_e) &=& \frac{A(T_i,T_i,T_e)}{A(T_i,T_j,T_e)} R(T_i,T_i,T_e) - \frac{A(T_i,T_i,T_j)}{A(T_i,T_j,T_e)} R(T_i,T_i,T_j).
\end{eqnarray}
When integrating the swaption payoff along the joint distribution, the distribution $pdf_Y$ corresponding to $R(T_i,T_i,T_j)$ will be integrated out as the payoff does not explicitly depend on it. Nevertheless, the impact of $R(T_i,T_i,T_j)$ will still be felt through the correlation. There are two sources of the correlation in this model. The correlation $\rho_{X_i,Y}$ between $R(T_i,T_i,T_e)$ and $R(T_i,T_i,T_j)$, which describes $R(T_i,T_j,T_e)$ as a joint density of 
$pdf_{X_i}$ and $pdf_Y$, and the correlation $\rho_{X_j}$  between $R(T_j,T_j,T_e)$ observed at $T_j$ with $R(T_i,T_j,T_e)$ observed at $T_i$. The latter  coincides with $\rho(T_i,T_j,R(T_j,T_e))$  as defined in Theorem~\ref{MRSCorr}. We can write the relationship in terms of independent Brownian motions $W^{X_i}_{T_i}$, $W^{Y}_{T_i}$, $W^{Z}_{T_i}$, $W^{X_j}_{T_i}$, and $W^{X_j}_{T_iT_j}$:
\begin{eqnarray}
X_i&=&(\cdots) + \tilde \sigma_{X_i} (\sqrt{1-\rho^2_{X_i,Y}}W^{X_i}_{T_i} + \rho_{X_i,Y}W^Y_{T_i}),\nonumber \\
Y&=&(\cdots) + \tilde \sigma_Y W^Y_{T_i},\nonumber \\
Z&=&(\cdots) + \tilde \sigma_Z W^Z_{T_i},\nonumber \\
X_j&=&(\cdots) + \tilde \sigma_{X_j} \left( W^{X_j}_{T_iT_j} + \rho_{X_j}W^Z_{T_i} +\sqrt{1-\rho^2_{X_j}}W^{X_j}_{T_i}\right),
\end{eqnarray}
where by $\tilde \sigma$ we underline that the volatilities do not include the square root time factor and by $(\cdots)$ we suppress the drifts (which do not impact correlations). 
 Equating (with omitted drifts)
\begin{eqnarray}
\tilde \sigma_Z W^Z_{T_i} &=&  \frac{A(T_i,T_i,T_e)}{A(T_i,T_j,T_e)} X_i - \frac{A(T_i,T_i,T_j)}{A(T_i,T_j,T_e)} Y\nonumber \\
&=& \frac{A(T_i,T_i,T_e)}{A(T_i,T_j,T_e)} \tilde \sigma_{X_i} \sqrt{1-\rho^2_{X_i,Y}}W^{X_i}_{T_i} +\left( \frac{A(T_i,T_i,T_e)}{A(T_i,T_j,T_e)}\tilde  \sigma_{X_i}\rho_{X_i,Y}  - \frac{A(T_i,T_i,T_j)}{A(T_i,T_j,T_e)}\tilde \sigma_Y\right)W^Y_{T_i},\nonumber\\
\end{eqnarray}
and assuming deterministic ratio of annuities, we derive the correlation between $X_i$ and $\E^{{\cal A}_n}[X_j|{\cal F}_{T_i}]$ as
\begin{eqnarray}
\rho_{X_i,X_j}&=&\left(\frac{A(t,T_i,T_e)}{A(t,T_j,T_e)} \frac{\sigma_{X_i}}{\sigma_Z} - \frac{A(t,T_i,T_j)}{A(t,T_j,T_e)}
\frac{ \sigma_Y}{ \sigma_Z}\rho_{X_i,Y}\right)\rho_{X_j}
\label{GCM}
\end{eqnarray}
($\tilde\sigma$ notation is no longer required as all the volatilities in the last formula have the same expiry).

\

We shall now consider two practical simplifications of the generic valuation for Bermudan swaptions described above.

\subsubsection*{Berms: Perfect Correlation of the Forward Rates}

\begin{Th}
\label{GMMC}
Under the assumption of the deterministic ratios of annuities and perfect correlations in all the forward rates,  Bermudan swaption payoff is
\begin{eqnarray}
\frac{B(t,\{T_j\},T_e,K)}{A(t,T_n,T_e)} &=&\E^{{\cal A}_n}\Bigl[max\Bigl(\frac{A(t,T_1,T_e)}{A(t,T_n,T_e)}(\omega(x_1-K))^+,\dots,
\frac{A(t,T_n,T_e)}{A(t,T_n,T_e)}(\omega(x_n-K))^+\Bigr)\nonumber\\
&&\Big|x_n=R(T_n,T_n,T_e),x_{n-1} = R(T_{n-1},T_{n-1},T_e)\dots, x_1=R(T_1,T_1,T_e),{\cal F}_t\Bigr],\nonumber\\
\label{BPC}
\end{eqnarray}
where the correlation matrix of the joint distribution is
\begin{eqnarray}
Corr&=&\left(
\begin{array}{c}
\rho_{ij}
\end{array}
\right),
\nonumber \\
\rho_{i<j} &=& \frac{A(t,T_i,T_e)^2\sigma^2_{X_i}+A(t,T_j,T_e)^2\sigma^2_{X_j} - A(t,T_i,T_j)^2\sigma^2_{R(T_i,T_i,T_j)}}{2A(t,T_i,T_e)A(t,T_i,T_j)\sigma_{X_j}\sigma_{R(T_i,T_i,T_j)}},\nonumber\\
\rho_{i>j} &= & \rho_{ji},\nonumber\\
\end{eqnarray}
and we assume  for any $i$ that $\rho_{ii}=1$. 
\end{Th}
\noindent
{\bf Proof:} If we work under the assumption of deterministic ratios of annuities,  then we can replace each $\frac{A(T_i,T_i,T_e)}{A(T_i,T_n,T_e)}$ in~(\ref{GENERICE}) by $\frac{A(t,T_i,T_e)}{A(t,T_n,T_e)}$, use pdfs of the individual swap rates in their own annuity measures  and evaluate the payoff as~(\ref{BPC}). 
For the correlation matrix we can simplify~(\ref{GCM}) because $\rho_{X_j}=1$, and for any $1\le i<j\le n$
\begin{eqnarray}
\rho_{X_i,X_j}&=&\frac{A(t,T_i,T_e)}{A(t,T_j,T_e)} \frac{\sigma_{X_i}}{\sigma_{X_j}} - \frac{A(t,T_i,T_j)}{A(t,T_j,T_e)}
\frac{\sigma_{R(T_i,T_i,T_j)}}{\sigma_{X_j}}\rho_{R(T_i,T_i,T_e),R(T_i,T_i,T_j)}\nonumber\\
&=&\frac{A(t,T_i,T_e)}{A(t,T_j,T_e)} \frac{\sigma_{X_i}}{\sigma_{X_j}} - \frac{A(t,T_i,T_j)}{A(t,T_j,T_e)} \frac{\sigma_{R(T_i,T_i,T_j)}}{\sigma_{X_j}}\times\nonumber\\
&&\frac{A(t,T_i,T_e)^2\sigma^2_{X_i}+A(t,T_i,T_j)^2\sigma^2_{R(T_i,T_i,T_j)}-A(t,T_j,T_e)^2\sigma^2_{X_j}}{2A(t,T_i,T_e)A(t,T_i,T_j)\sigma_{X_i}\sigma_{R(T_i,T_i,T_j)}}\nonumber\\
&=&\frac{A(t,T_i,T_e)^2\sigma^2_{X_i}+A(t,T_j,T_e)^2\sigma^2_{X_j} - A(t,T_i,T_j)^2\sigma^2_{R(T_i,T_i,T_j)}}{2A(t,T_i,T_e)A(t,T_i,T_j)\sigma_{X_j}\sigma_{R(T_i,T_i,T_j)}}.\nonumber\\
\end{eqnarray}

\subsubsection*{Berms: Lattice Pricing }
A richer model can be obtained in the lattice framework similar to Hagan LGM~\cite{H}.
 Assume that at the time $T_j$ we know the conditional expectation
\begin{eqnarray}
f^{(j)}(x_j) & =& \E^{{\cal_A}_j}\Bigl[\frac{B(T_j)}{A(T_j,T_j,T_e)}\Bigl|R(T_j,T_j,T_e),{\cal F}_{T_j}\Bigr]
\end{eqnarray} 
as a function $f^{(j)}(\cdot )$ of $x_j=\E^{{\cal A}_j}[R(T_j,T_j,T_e)| {\cal F}_{T_j}]$,
where ${\cal A}_j$ is the annuity measure corresponding to the $j$th coterminal swap $S(T_j,T_e,K)$ and by the superscript $(j)$ we express that the only part of the swaption active is the one to be exercised in to coterminals starting from time $T_j$ and beyond.
\begin{Th}
\label{GHL}
 The Bermudan payoff roll back formula is:
\begin{eqnarray}
f^{(j-1)}(x_{j-1})&=& \frac{A(t,T_{j},T_e)}{A(t,T_{j-1},T_e)}\E^{{\cal_A}_{j}}\Bigl[\E^{{\cal_A}_j}\Bigl[max\Bigl(\frac{A(T_{j-1},T_{j-1},T_e)}{A(T_{j-1},T_{j},T_e)}(\omega(x_{j-1}-K))^+,\E^{{\cal_A}_j}\Bigl[f^{(j)}(z+\tilde x_j)|{\cal F}_{T_{j-1}}\Bigr]\Bigr)\nonumber\\
&&\Bigl|\tilde x_j=\E^{{\cal_A}_{j}}[R(T_j,T_j,T_e)|{\cal F}_{T_{j-1}}],{\cal F}_{T_{j-1}}\Bigr ]\Bigl |x_{j-1} = R(T_{j-1},T_{j-1},T_e),{\cal F}_{T_{j-1}}\Bigr],
\label{LTC}
\end{eqnarray}
where
\begin{eqnarray}
z&=&\E^{{\cal_A}_j}\Bigl[R(T_j,T_{j},T_e) - \E^{{\cal_A}_j}[R(T_j,T_{j},T_e)|{\cal F}_{T_{j-1}}]\Bigl | {\cal F}_{T_j}\Bigr].
\end{eqnarray}
\end{Th}
\noindent
{\bf Proof:} Let $W^{(j)}(t)$, $j=1,\dots,n$,  be the European swaption which can be exercised in to the coterminal swap $S(T_j,T_e,K)$ of the Bermudan swaption $B(t)$ at its expiry date $T_j$. We proceed as
\begin{eqnarray}
f^{(j-1)}(x_{j-1}) & =&
\E^{{\cal_A}_{j-1}}\Bigl[\frac{B(T_{j-1})}{A(T_{j-1},T_{j-1},T_e)}\Bigl |R(T_{j-1},T_{j-1},T_e),{\cal F}_{T_{j-1}}\Bigr]\nonumber\\
& =&\frac{A(t,T_{j},T_e)}{A(t,T_{j-1},T_e)}\E^{{\cal_A}_{j}}\Bigl[\frac{B(T_{j-1})}{A(T_{j-1},T_{j},T_e)}\Bigl |R(T_{j-1},T_{j-1},T_e),{\cal F}_{T_{j-1}}\Bigr]= \frac{A(t,T_{j},T_e)}{A(t,T_{j-1},T_e)} (*).\nonumber\\
\end{eqnarray}
\begin{eqnarray}
(*)&=&\E^{{\cal_A}_{j}}\Bigl [max\left(\frac{A(T_{j-1},T_{j-1},T_e)}{A(T_{j-1},T_j,T_e)}\frac{W^{(j-1)}(T_{j-1})}{A(T_{j-1},T_{j-1},T_e)},\E^{{\cal_A}_{j}}\Bigl[\frac{B(T_j)}{A(T_j,T_{j},T_e)}\Bigl |{\cal F}_{T_{j-1}}\Bigr]\right)\Bigl |R(T_{j-1},T_{j-1},T_e),{\cal F}_{T_{j-1}}\Bigr ]\nonumber\\
&=&\E^{{\cal_A}_{j}}\Bigl[max\Bigl(\frac{A(T_{j-1},T_{j-1},T_e)}{A(T_{j-1},T_{j},T_e)}(\omega(x_{j-1}-K))^+,\E^{{\cal_A}_j}\Bigl[\E^{{\cal_A}_{j}}\Bigl[f^{(j)}\Bigl (x_j\Bigr )\Bigl |x_{j} =R(T_j,T_{j},T_e),{\cal F}_{T_j}\Bigr]\Big| {\cal F}_{T_{j-1}}\Bigr]\Bigr)\nonumber\\
&&\Big|x_{j-1} =R(T_{j-1},T_{j-1},T_e),{\cal F}_{T_{j-1}}\Bigr]\nonumber\\
&=&\E^{{\cal_A}_{j}}\Bigl[max\Bigl(\frac{A(T_{j-1},T_{j-1},T_e)}{A(T_{j-1},T_{j},T_e)}(\omega(x_{j-1}-K))^+,\E^{{\cal_A}_{j}}\Bigl[f^{(j)}\Bigl (z+\E^{{\cal_A}_j}[R(T_j,T_{j},T_e)|{\cal F}_{T_{j-1}}]\Bigr )\Bigl |{\cal F}_{T_{j-1}}\Bigr]\Bigr)\nonumber\\
&&\Bigl |x_{j-1} =R(T_{j-1},T_{j-1},T_e),{\cal F}_{T_{j-1}}\Bigr]\nonumber\\
&=&\E^{{\cal_A}_{j}}\Bigl[\E^{{\cal_A}_j}\Bigl[max\Bigl(\frac{A(T_{j-1},T_{j-1},T_e)}{A(T_{j-1},T_{j},T_e)}(\omega(x_{j-1}-K))^+,\E^{{\cal_A}_j}\Bigl[f^{(j)}(z+\tilde x_j)|{\cal F}_{T_{j-1}}\Bigr]\Bigr)\nonumber\\
&&\Bigl|\tilde x_j=\E^{{\cal_A}_{j}}[R(T_j,T_j,T_e),{\cal F}_{T_{j-1}}]\Bigr ]\Bigl |x_{j-1} = R(T_{j-1},T_{j-1},T_e),{\cal F}_{T_{j-1}}\Bigr].
\end{eqnarray}
This  coincides with the right hand side of~(\ref{LTC}) up to the scalar $\frac{A(t,T_{j},T_e)}{A(t,T_{j-1},T_e)}$.

\
 
The payoff~(\ref{LTC}) can be evaluated by the same process as in the generic case of the Canary swaption from the previous section. The overall payoff of a Bermudan trade can be evaluated  stepping backwards through all the swaption expiries as in~\cite{H}. 

\

We performed a comparison of valuations based on Theorem~\ref{GMMC} and Theorem~\ref{GHL} in real market conditions observed close of business day on 30 June 2025 against the classical Hagan LGM. The volatility surface used for all three models is SABR surface calibrated to the observed European swaption market. The maximum in Theorem~\ref{GMMC} is evaluated  by the moment matching techniques as outlined in~\cite{SZ}. A direct application of Theorem~\ref{GMMC} results in valuations which are significantly below the market. To demonstrate the qualitative behaviour of the corresponding model we adjusted the correlation close to 100\%.  The valuation based on Theorem~\ref{GHL} is performed using a) Hagan lattice step for the last inside expectation $\E^{{\cal_A}_j}\Bigl[f^{(j)}(z+\tilde x_j)|{\cal F}_{T_{j-1}}\Bigr]$ and b)  Simpson rule  integration for the overall conditional  expectation (the middle expectation is used to express the functional dependence on two variables: $x_{j-1}$ and $\tilde x_j=\E^{{\cal_A}_j}[R(T_j,T_j,T_e)|{\cal F}_{T_{j-1}}]$). The overall conditional expectation is a function of $x_{j-1}$ and is re-used at the next step of Hagan lattice scheme. We use a linear term structure ranging from  88\% to 99\%  for the correlations between the coterminal swap rates and 1Y swap rates. The moments matching is based on 99.9\% correlations. Figure 2 shows the difference of each of three valuations with the market consensus for a range of strikes between ATM-300bps and ATM+600bps available for a payer callable swap which starts in 5 years, last 10 years and has the annual exercise. The doted lines are plus/minus one standard deviations of the individual market quotes.

\begin{figure}[h!]
\centering
\includegraphics[width=0.8\textwidth]{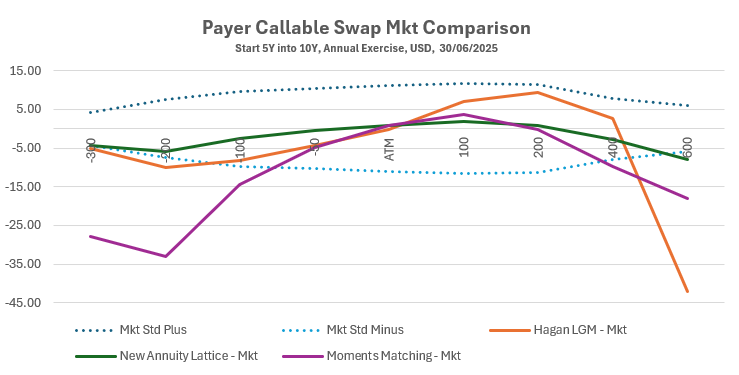}
\small{\footnotesize\caption{Model price vs market consensus under different model assumptions: the callable swap equivalent to the Bermudan swaption on 10Y swap starting in 5Y with the annual exercise frequency.}}
\label{PC5Y1Y10Y}
\end{figure}

\section{Stochastic Ratios of Annuities}

While Theorem~\ref{GHL} and its derivation do not depend on the deterministic annuity ratio assumption, the numerical implementation does require to choose the dynamics for the ratios. 
For the numerical experiment in the previous section we assumed that the annuity ratios are deterministic and the Hagan  lattice style model  showed a good fit to the market. 

The deterministic annuity ratio assumption is widely used in financial engineering (see, for example, \cite{AP3}). It relies on empirical observation that an annuity ratio has a low variance and is an approximately slow changing linear function of the corresponding long and short swap rates. Nevertheless, effects like mild correlation skews in midcurve swaptions and CMS spread options markets make it beneficial  to have a handle on the stochastic part of annuities ratios.

For a Bermudan swaption to reflect the dependence of the price on stochastic ratios of annuities as an alternative to the correlation skew management we suggest using Theorem~\ref{TLNZ} from  Section 1. The convexity adjustments of the long and short swap rates will translate into the corresponding convexity adjustment of the forward starting swap rate (see the conversion formulas in~\cite{KF}). Therefore, a single distribution shift/convexity adjustment parameter per each coterminal swap rate distribution would suffice to control the stochastic ratios at least in the first order. 

\section*{Conclusion}
We developed an alternative model for Bermudan swaption valuations which allows us to express explicitly the dependence of Bermudan swaption price on the swap rate distributions and correlations between them. The model replaces the use of mean reversion parameters (as in traditional approaches) with swap rate correlations, which can be estimated from historical data or implied from midcurve swaption and CMS spread markets. The model does not require product specific calibrations, and its lattice style analogy has only that many steps as the number of swaption exercises. The approach offers potential to a faster and more accurate risk management of  Bermudan swaptions.
We provided a numerical implementation based on the implied volatility-by-strike swap rate distribution parameterisation and the deterministic annuity ratio assumption. The model demonstrated a good fit to the market.

\

\noindent
{\bf Acknowledgement.} The author would like to express a sincere gratitude to Andrew Green, Shuqing Ma and Scotiabank Innovation Hub for supporting this research. I am also grateful to various  Scotiabank GAFE team members for creating a friendly research environment and for useful discussion with Chris Dennis, Bin Guo,  Michael Konikov, Dirk Petera, Duan Qian, Rex Sutton and Alex Zemlianov.

\end{document}